\documentclass[graybox]{svmult}

\usepackage{type1cm}        
\usepackage{makeidx}         
\usepackage{graphicx}        
\usepackage{multicol}        
\usepackage[bottom]{footmisc}

\usepackage{newtxtext}       %
\usepackage[varvw]{newtxmath}       
\usepackage{bm}
\usepackage{cite}
\newcommand{\ds}{\displaystyle}

\renewcommand{\leq}{\leqslant}
\renewcommand{\geq}{\geqslant}

\newcommand{\xx}{\mathbf{x}}

\newcommand{\R}{\mathbb{R}}

\newcommand{\F}{\mathcal{F}}

\newcommand{\y}{{Y}}

\newcommand{\equilibrium}{E} 
\newcommand{\ftrunc}[1]{f_{({#1})}}
\newcommand{\ftruncE}[1]{\ftrunc{N}^\equilibrium}

\newcommand{\inta}{\int_{\mathbb{R}^3} \int_0^\infty }

\makeindex             

\begin{document}

\title*{The Maximum Entropy Principle in Nonequilibrium Thermodynamics: A Brief History and the Contributions of Wolfgang Dreyer }
\titlerunning{Maximum Entropy Principle in Nonequilibrium Thermodynamics}

\author{Takashi Arima and Tommaso Ruggeri\\ \vspace{0.5cm}
{\small Dedication:  This paper is dedicated to the memory of Wolfgang Dreyer. In particular, one of the authors, T.R., was a close friend and colleague of Wolfgang and always greatly appreciated his outstanding contributions as a scientist.} }

\authorrunning{T. Arima and T. Ruggeri}
\institute{
Takashi Arima \at Department of Engineering for Innovation. National Institute of Technology, \\ Tomakomai College, Tomakomai, {Japan}\\
\email{arima@tomakomai-ct.ac.jp}\and
Tommaso Ruggeri  \at  Department of Mathematics, University of Bologna,  and Accademia dei Lincei,  Italy \\ \email{tommaso.ruggeri@unibo.it}}
%
%

\maketitle

\abstract{- We present a brief history of how the famous Maximum Entropy Principle was used as closure of moments of the Boltzmann equation. In particular, we want to remark on the important role of two fundamental papers by Wolfgang Dreyer, one in the classical framework and one in a relativistic context, to use this principle and to compare the result with the macroscopic theory of Rational Extended Thermodynamics.}

\section{Introduction}
The \emph{Maximum Entropy Principle} (MEP), rooted in probability theory and statistical mechanics, was formalized in its modern form by Edwin T. Jaynes in 1957 \cite{Jaynes1,Jaynes2}. Jaynes developed the principle to provide a general approach for statistical inference, extending beyond physics to other fields where systems might be out of equilibrium (see e.g., \cite{kapur}).
The MEP has served as both a philosophical and practical foundation for making unbiased, informed predictions in the presence of incomplete data. In image reconstruction, it helps create images that align with known data without introducing unwarranted assumptions. In modern AI, the MEP continues to influence techniques for uncertainty management, probabilistic inference, and decision-making, ensuring that models remain general and data-driven without making unnecessary assumptions. 

In the context of nonequilibrium thermodynamics, the MEP has become a powerful tool. It aids in analyzing and predicting the behavior of systems far from equilibrium, where conventional equilibrium assumptions do not apply. By maximizing entropy under given constraints, MEP offers a systematic and unbiased method to model complex, evolving, and fluctuating systems, making it widely applicable in physical, biological, and socio-economic domains. 

Regarding the closure of moments associated with the Boltzmann equation, the MEP is widely known as the Levermore procedure \cite{levermore}. However, significant work in this area predates this paper, notably conducted by Kogan \cite{kogan}, Dreyer \cite{dreyer1,Dreyer2}, and M\"uller and Ruggeri \cite{ET}. The aim of this paper is to provide a brief history of the contributions made on this subject, particularly emphasizing the crucial role that Wolfgang Dreyer played in both classical and relativistic frameworks. In particular, we also highlight Dreyer’s significant contributions in elucidating the equivalence between the MEP and the nonequilibrium thermodynamics framework of Rational Extended Thermodynamics (RET) \cite{ET,RET,beyond,newbook}.

The paper is organized as follows: In Section 2, we present the history of MEP in the classical (non-relativistic) context, and in Section 3, we examine it in the relativistic framework. 

\section{MEP in the Kinetic Theory of Gases and Connection with Rational Extended Thermodynamics: Classical Framework}
In this section, we present a concise history of the MEP as a closure procedure for the moment equations derived from the classical Boltzmann equation, as documented in the literature available to the authors. This discussion expands upon the mathematical details provided in the recent book by Ruggeri and Sugiyama \cite{newbook}.

The initial applications of MEP were focused on monatomic gases, establishing the foundational framework.
\subsection{Moment Equations and Problem of Closure}
The kinetic theory of gases describes the state of a rarefied  monatomic gas using the velocity distribution function \( f(t, \mathbf{x}, \boldsymbol{\xi}) \). Here, \( f(t, \mathbf{x}, \boldsymbol{\xi}) \, d\mathbf{x} \, d\boldsymbol{\xi} \) denotes the number of particles within the phase-space volume element \( d\mathbf{x} \, d\boldsymbol{\xi} \) at position \( (\mathbf{x}, \boldsymbol{\xi}) \in \mathbb{R}^3 \times \mathbb{R}^3 \). In the absence of external forces, the evolution of \( f \) is governed by the Boltzmann equation:
\begin{equation}
    \partial_t f + \xi_i \partial_i f = Q(f), \label{Boltzmann}
\end{equation}
where the collision integral \( Q(f) \) encapsulates the rate of change in \( f \) due to molecular collisions. Here, \( \partial_t \) and \( \partial_j \) denote partial derivatives with respect to time \( t \) and spatial coordinate \( x_j \) \ $(j=1,2,3)$, respectively. Summation over repeated indices follows Einstein's convention unless explicitly stated otherwise.

The distribution function allows us to construct macroscopic quantities defined as velocity-space moments of \( f \):
\begin{equation}\label{momenti}
  F = m \int_{\mathbb{R}^3} f \, d\boldsymbol{\xi}, \quad F_{i_1 \dots i_n} = m \int_{\mathbb{R}^3} f \, \xi_{i_1} \cdots \xi_{i_n} \, d\boldsymbol{\xi}, \quad i_k \in \{1, 2, 3\}.
\end{equation}
From  the Boltzmann equation, the governing equations for these moments are expressed as a set of infinite balance laws\footnote{For \( n=0 \), we assume \( F_{i_1 i_2 \dots i_n} = F \), \( F_{i_1 i_2 \dots i_n j} = F_j \) and \( P_{i_1 i_2 \dots i_n} =P= 0 \).}:
\begin{equation}
    \partial_t  F_{i_1 \dots i_n}  + \partial_j  F_{i_1 \dots i_n j} =  P_{i_1 \dots i_n}, \qquad n=0,1,\dots \label{moments}
\end{equation}
where  
\begin{align*}
&  F_{i_1 \dots i_n j} =m \int_{\mathbb{R}^3}  f\, \xi_j \xi_{i_1} \cdots \xi_{i_n}  \, d\boldsymbol{\xi}, \qquad 
P_{i_1 \dots i_n} =m \int_{\mathbb{R}^3}  Q(f)\,  \xi_{i_1} \cdots \xi_{i_n}  \, d\boldsymbol{\xi}.
\end{align*}
Since the first five equations for $F$, $F_i$, and $F_{ll}$ represent the conservation laws of mass, momentum, and energy, by comparing these laws of continuum mechanics, we can deduce the physical meanings of the first 13 moments: $F$ represents the mass density, $F_i$ the momentum density, $F_{ll}$ twice the total energy density, $F_{ij}$ the momentum flux, and $F_{lli}$ twice the energy flux.
In particular, the first five production terms, $P$, $P_i$, and $P_{ll}$, vanish due to the conservation laws.

To truncate this hierarchy at tensorial order \( N \), we need to know the expression of the last flux $F_{i_1 i_2 \dots i_{N+1}}$ and of the production terms $P_{i_1 \dots i_n}$ with $n=2,\dots N$ and $P_{ll}=0$. The truncated balance laws construct the following system
\begin{equation} \label{eq:hierarchy-truncated}
	\partial_t F_A + \partial_i F_{iA} = P_A, \qquad  0 \leq A \leq N,
\end{equation}
where the following notations are introduced for the sake of compactness:
\begin{equation}\label{FAA}
	F_A = \left\{
	\begin{array}{lll}
		F &  & \text{for } A=0 \\
		F_{i_1 \cdots i_A} &  & \text{for } 1\leq A\leq N
	\end{array}\right., \qquad 
	F_{iA} = \left\{
	\begin{array}{lll}
		F_i &  & \text{for } A=0 \\
		F_{i\,i_1 \cdots i_A} &  & \text{for } 1\leq A\leq N
	\end{array} \right.,
\end{equation}
\begin{equation*}
	 P_A = \left\{
	\begin{array}{lll}
	 0 & & \text{for } A=0,1 \\
		P_{i_1 \cdots i_A} &  & \text{for } 2\leq A\leq N \ \ \  (\text{with  }P_{ll}=0).
	\end{array} \right.
\end{equation*}
With these symbols we have:
\begin{equation*}
    F_A = m \int_{\mathbb{R}^3}  f\, \xi_A \, d\boldsymbol{\xi}, \quad  
    F_{iA} = m \int_{\mathbb{R}^3}  f\, \xi_i \xi_A \, d\boldsymbol{\xi}, \quad  
    P_A = m \int_{\mathbb{R}^3}  Q\, \xi_A \, d\boldsymbol{\xi},
\end{equation*}
where 
\begin{equation*}
    \xi_A = \left\{
	\begin{array}{lll}
		1 &  & \text{for } A=0 \\
		\xi_{i_1} \cdots \xi_{i_A} &  & \text{for } 1\leq A\leq N .
	\end{array}\right. 
\end{equation*}

\subsection{Kogan’s MEP for Non-Degenerate Gases (1967)}\label{koganino}
In 1967, Mikhail N. Kogan from the Academy of Sciences of the USSR, Moscow, authored a book in Russian, which was translated in 1969 under the title \emph{Rarefied Gas Dynamics} \cite{kogan}. This book offers an elegant presentation of the Boltzmann equations. In Chapter III, \emph{General Methods of Solution of the Boltzmann Equation}, Kogan introduces the method of moments. Notably, in Section 3.16, page 267, there is a section titled \emph{The Maximum Probability Principle}. Here, he begins by noting that, in equilibrium, the Maxwellian distribution maximizes the probability under the constraint that the first five moments, i.e., $F$, $F_i$, and $F_{ll}$, are prescribed. He then writes:

\begin{quote} It is natural to try to find the distribution function as the most probable function also for nonequilibrium processes. 
\end{quote}
For this aim, he considered non-degenerate gas, of which entropy density is given by
\begin{align*}
    h &= -k_B \int_{\mathbb{R}^3} f  \ln  \frac{f}{y}   \, d\boldsymbol{\xi}, 
\end{align*}
being $k_B$ the Boltzmann constant and $y$ is a constant with the same dimension of $f$ that  is defined as
\begin{equation*}
    y=e(2s +1) \frac{m^3}{h^3}
\end{equation*}
where  \( h \) is Planck's constant, and $s$ is the spin quantum number of a particle \footnote{In Kogan's book \cite{kogan}, the value $y=1$ is chosen. In fact, in many references, $y=1$ or $y=e$ is often used.}.

Then, Kogan required that the entropy is maximal under the constraints that the moments \eqref{momenti} are prescribed until the tensorial index $N$: $n=0,\dots,N$.
Therefore, he constructed the functional
\begin{equation*}
	\mathcal{L} \left(f\right) = - k_B   \int_{\R^3}  f  \ln  \frac{f}{y}\, d\boldsymbol{\xi} +
\Lambda_A \left(F_A - m \int_{\mathbb{R}^3} f \, \xi_A\, d\boldsymbol{\xi}\right),
\end{equation*}
where $\Lambda_A \equiv \Lambda_A(t, \mathbf{x})$, defined in a similar way to $F_A$ in \eqref{FAA}, are the Lagrange multipliers, and we omit the summation over $A$ from $0$ to $N$.
Requiring the constraint maximum, he found immediately that the distribution function such that $\delta \mathcal{L} \left(f\right) =0$,  $\delta^2 \mathcal{L} \left(f\right)<0$, is
\begin{equation}\label{bravoKogan}
    f_N = y \exp\left(-1 - \frac{m}{k_B}\chi\right), \qquad \chi = \Lambda_A \xi_A. 
\end{equation}
\emph{Therefore, Kogan was the first to derive the expression for the distribution function that maximizes entropy for non-degenerate gases under the constraint that a specified number of moments are prescribed. }

By inserting  \eqref{bravoKogan} into the last flux and into the production terms, he was able to obtain a closed system having as unknown the Lagrange multipliers. Kogan was also the first to observe that this approach encounters issues related to the integrability of moments with $f_N$. He wrote:
\begin{quote}
    However, formal use of the method immediately encounters several well-known difficulties.
    
The first difficulty is of a mathematical nature. Suppose that we examine a problem in which the highest moment determining the problem is an odd moment, for example, the heat flux $q_i$. Then, $\dots$ we obtain divergent integrals $\dots$  and the problem has no solution. 
\end{quote}
\emph{Therefore, again, Kogan was the first to realize that in the nonlinear closure, it is necessary (but not sufficient) that  $N$ is even such that the moments are integrable}.  

In particular, the 13-moment model which corresponds to a specific instance of 
$N =3$ and only the trace part in the third order tensor is taken into account,  has no significance in the MEP.

\emph{Additionally, he was also the first to consider the approximation near equilibrium}. He wrote on page 271\footnote{Here, we have modified the equation number (originally Eq.~(16.5) in \cite{kogan}) and the mathematical expressions in \eqref{appro} to correspond to the context and notation used in this paper.}:
\begin{quote}
 Therefore, we may speak of the distribution function as most probable only for flows with small Knudsen numbers, i.e., for flows close to local equilibrium. In that case, all the moments, with the exception of the hydrodynamic ones, may be considered small, and we may linearize expression \eqref{bravoKogan} relative to the equilibrium distribution function, i.e., we may write it in the form
\end{quote}
\begin{equation}\label{appro}
    f_N = f_E \left( 1 - \frac{m }{k_B} \tilde{\chi} \right), \qquad \tilde{\chi} = 
    \tilde{\Lambda}_A \xi_A.
\end{equation}
Here, the $\tilde{\Lambda}$'s represent the difference between the nonequilibrium and equilibrium Lagrange multipliers, and $f_E$ is the Maxwellian distribution. As the Maxwellian is given by a convergent exponential that dominates any polynomial, the moments are always integrable for any $N$. 

Kogan, then, considered $13$ moments by using the approximated distribution function \eqref{appro} and demonstrated its equivalence to the Grad distribution function \cite{Grad}. Therefore, the Grad system also maximizes entropy, but only near the equilibrium state. On the other hand, Grad's procedure is a perturbative method relative to equilibrium.

\subsection{Dreyer’s Work and Comparison between MEP and RET (1987)}
Twenty years after Kogan's works, Wolfgang Dreyer who was previously a student of Ingo M\"uller, first presented fundamental results concerning MEP  in both the classical \cite{dreyer1} and relativistic frameworks \cite{Dreyer2}. Here, we exhibit his contribution to the classical context. Later, in section \ref{sec:rel}, we discuss his contribution to the relativistic framework.

Dreyer generalized Kogan’s result for the first time, presenting the MEP not only for non-degenerate gases but also for degenerate ones, whose entropy is given by
\begin{align}
h = -k_B \int_{\mathbb{R}^3} \left\{ \left(s^2 -1 + \ln \frac{f}{\y}\right) + s \frac{\y}{f} \left( 1 - s \frac{f}{\y}\right) \ln \left( 1 -s \frac{f}{\y} \right) \right\} f \, \, d\boldsymbol{\xi},
\label{hgen}
\end{align}
where $\y=y/e$.  The constant  $s$ is $0$ for non-degenerate gases and $s=1$ for Fermions, and $-1$ for Bosons.
For this entropy, the $f_N$ that maximizes the entropy is:
\begin{equation}\label{tom}
  f_N =  \frac{\y}{e^{\frac{m }{k_B}\chi} + s }.
\end{equation}
For $s=0$, \eqref{tom} reduce to the Kogan's one \eqref{bravoKogan}, while for degenerate gases becomes:
\begin{align}\label{Drey}
f_N =  \frac{\y}{e^{\frac{m}{k_B}\chi} \mp 1},
\end{align}
where $\chi$ is given by \eqref{bravoKogan}$_2$, and the upper sign is for Bosons, and the lower sign is for Fermions.

\emph{Dreyer's closure with $f_N$ given in \eqref{Drey}  is fully nonlinear, as it does not rely on an expansion near equilibrium, and for non-degenerate gases, he was the first to use MEP to close the system of moments.}

Nevertheless, the main aim of Dreyer \cite{dreyer1} was to prove the equivalence between the closure using  MEP and the one at the phenomenological level of  RET for rarefied gases, particularly in the $13$-moment case. Since RET was initially closed only near equilibrium, Dreyer considered a similar expansion to \eqref{appro}.

Before discussing the details, it is necessary to exhibit a brief survey about the  RET theory.

\subsubsection{Brief Introduction of Rational Extended  Thermodynamics}
 The first hyperbolic phenomenological model for a viscous, heat-conducting gas was introduced by Ingo M\"uller in his doctoral thesis \cite{mueller1}. It was based on modifying the Gibbs equation, recognizing that the hypothesis of so-called \emph{local equilibrium} is overly restrictive for phenomena far from equilibrium.  

However, Ruggeri \cite{Ruggeri-Acta} criticized this approach for several mathematical reasons, particularly because the differential system is not, a priori, in the form of a balance laws. This limitation implies that it is not possible to define weak solutions, making the study of shock waves impossible. Ruggeri, using the theory of hyperbolic systems compatible with an entropy law that can be symmetrized, proposed a model that is symmetric hyperbolic for any fields. It satisfies both the entropy principle and the convexity condition of entropy. However, the system's structure differs from that of moments. Consequently, M\"uller, along with Liu, revised Extended Thermodynamics, adopting the balance law structure dictated by the moment theory associated with the Boltzmann equation as a starting point \cite{LM}. In a relativistic context, a similar theory was developed by Liu, M\"uller, and Ruggeri \cite{LMR}.

This approach became known as \emph{Rational Extended Thermodynamics} and the main results were summarized in the books by M\"uller and Ruggeri \cite{ET, RET}.

The central idea of the RET approach is to use the structure of balance laws for moments truncated at order \( N \) (see \eqref{eq:hierarchy-truncated}). However, it then departs from viewing the   \( (F_A, F_{iA}, P_A) \)  as moments of a distribution function solution of the Boltzmann equation. Instead, the system is treated as a phenomenological model typical of continuum mechanics. As is common in such cases, the system involves more fields than equations, requiring constitutive equations to close the system.

In this context, the fields \( F_A \), representing densities, are treated as the unknowns, while the last flux term and production terms serve as the constitutive equations. The requirement is that these constitutive equations be local in nature:
\begin{align*}
\begin{split}
F_{{N+1}} \equiv F_{{N+1}}(F_B), \quad  
     P_{A} \equiv P_{A}(F_B), \quad (A,B = 0,\dots, N). 
\end{split}
\end{align*}

To determine the closure procedure, RET adopts the following universal principles:  
\begin{enumerate}  
    \item[(i)] \textit{Objectivity and relativity principles}: Constitutive equations must be objective, and balance laws must be invariant under Galilean transformations (or Lorentz transformations for relativistic gases).  
    \item[(ii)] \textit{Entropy principle}: This principle requires that any solutions of the balance laws \eqref{moments} satisfy an additional balance  law for an additive, objective scalar \(h\), called entropy:  
    \begin{equation} \label{EP}  
       \partial_t h + \partial_i h^i = \Sigma,  
    \end{equation}  
    where \(h_i\) and \(\Sigma\) are the entropy flux and entropy production, respectively. These are assumed to be local functions of the densities field:  
    \begin{equation*}  
        h = h(F_A), \quad h^i = h^i(F_A), \quad \Sigma = \Sigma(F_A).  
    \end{equation*}  
    Additionally, the entropy production must be non-negative: $ \Sigma \geq 0. $ 
\end{enumerate}  
The requirement that a system of balance laws  
\begin{equation}  \label{sistema}
\partial_t \mathbf{u} + \partial_i \mathbf{F}^i = \mathbf{P}  
\end{equation}  
(such as the moment equations \eqref{eq:hierarchy-truncated}) is compatible with an additional entropy balance law \eqref{EP} falls under the well-known problem of hyperbolic systems. Starting with Boillat's paper \cite{boillat1}, Ruggeri and Strumia \cite{RS} proved that if the entropy density \(h\) is a convex function of   \(\mathbf{u} \), there exists a privileged field called the \emph{main field} \(\mathbf{u}^\prime\) and four potentials 
\begin{align}
 h^\prime = \mathbf{u}^\prime\cdot  \mathbf{u} -h, \quad h^{\prime i} = \mathbf{u}^\prime\cdot  \mathbf{F}^i -h^i,    
 \label{4pot}
\end{align}
such that
\begin{align}\label{potenziali}
   \mathbf{u} = \frac{\partial h^\prime}{\partial \mathbf{u}^\prime}, \qquad  \mathbf{F}^i = \frac{\partial h^{\prime i}}{\partial \mathbf{u}^\prime},   
\end{align}
and inserting \eqref{potenziali} into the system \eqref{sistema} 
 assumes the structure of a special symmetric hyperbolic system, often referred to as Godunov type\footnote{Godunov in \cite{godunov} was the first to prove that the Euler fluid equations and systems derived from a variational principle can be written in the symmetric form \eqref{18}.}:  
\begin{equation}  \label{18}
   \frac{\partial^2 h^\prime}{\partial \mathbf{u}^\prime \partial \mathbf{u}^\prime  } \partial_t \mathbf{u}^\prime+ \frac{\partial^2 h^{\prime i}}{\partial \mathbf{u}^\prime \partial \mathbf{u}^\prime  } \partial_i \mathbf{u}^\prime = \mathbf{P}.
\end{equation}  
The matrix before the time derivative is positive-definite as $h^\prime$ is the Legendre transformation of $h$ and therefore a convex function of the dual field $\mathbf{u}^\prime$.

The main field is a particular set of Lagrange multipliers proposed by Liu \cite{ISL}, satisfying:  
\begin{align*}  
    \partial_t h + \partial_i h^i - \Sigma = \mathbf{u}^\prime \cdot \left( \partial_t \mathbf{u} + \partial_i \mathbf{F}^{i} - \mathbf{P} \right).
\end{align*}  
For more details, readers can refer also to the recent book by Ruggeri and Sugiyama \cite{newbook}.  
\subsubsection{Equivalence of Closure using the RET Principles  and the  MEP for Moment Systems}  
\emph{Dreyer was the first to prove in \cite{dreyer1} the equivalence between the Lagrange multipliers of MEP and the main field of the phenomenological theory compatible with the entropy principle.}

Specifically, in the 13-moment case near equilibrium, Dreyer demonstrated that MEP provides the same closure as phenomenological RET, thereby showing that three different closure methods—Grad's, RET's, and MEP's—yield the same system. This result was significant since the three methods are fundamentally different. Grad’s method uses a perturbation of the Maxwellian with Hermite polynomials, MEP uses maximum entropy at the kinetic level, while RET is a continuum theory that shares with kinetic theory only the structure of balance laws, where the next density is the previous flux, and the first five equations are the conservation laws for mass, momentum, and energy.

\subsection{M\"uller and Ruggeri’s Higher-Order Moment Systems and Symmetric Hyperbolic Systems (1993) } \label{MURU}
In 1993, Clifford Truesdell invited Ingo M\"uller and Tommaso Ruggeri to publish a book \cite{ET} on the recent developments in the theory of \emph{Extended Thermodynamics} as part of the series he edited, \emph{Springer Tracts in Natural Philosophy}. This emerging theory also aimed to resolve the longstanding debate that arose in the literature following Müller's landmark paper \cite{mullerfourier}, where he demonstrated that the Fourier and Navier-Stokes "constitutive" equations violate the principle of objectivity (frame indifference). 

At the time, Müller believed his findings suggested that the objectivity principle was not universally valid. This sparked extensive debate, with scholars dividing into proponents and opponents of the objectivity principle.

Independent contributions by Bressan \cite{bressan} and Ruggeri \cite{ruggeri_padova} offered a different perspective. They proposed that M\"uller's result did not undermine the objectivity principle itself but instead indicated that the Fourier and Navier-Stokes equations were not genuine constitutive relations. 

The precise and compelling resolution came from RET. In RET, in fact, the Navier-Stokes and Fourier laws are shown to be approximations derived from the balance laws of 13 moments, valid only when certain relaxation times are small. Consequently, these laws are not true constitutive equations and do not need to adhere to the objectivity principle at the same level as fundamental balance laws like momentum and energy, which must only satisfy Galilean invariance\footnote{Some authors hold a different opinion and continue to consider simplified hyperbolized formulations of Fourier's law and Navier-Stokes equations, such as the Cattaneo equation or Maxwell-type viscoelasticity models, as constitutive equations. To restore the objectivity, they introduce additional derivatives into these models.}.

Returning to the historical contribution on the MEP in Chapter 9 of the first edition of their book in 1993 \cite{ET}, M\"uller and Ruggeri, considering non-degenerate gases, highlighted the clear equivalence between MEP and  RET closure. Moreover they showed that the moments \eqref{momenti}, derived from the distribution function \eqref{bravoKogan} for the truncated system, satisfy the following symmetric relations (see p. 171 of \cite{ET}):  
\begin{align*}  
    \frac{\partial F_A}{\partial \Lambda_B} = \frac{\partial F_B}{\partial \Lambda_A}, \quad  
    \frac{\partial F_{iA}}{\partial \Lambda_B} = \frac{\partial F_{iB}}{\partial \Lambda_A}.  
\end{align*}  
These relations imply the existence of potentials \eqref{4pot}, enabling the truncated moment system to be formulated as a symmetric hyperbolic system of the Godunov type \eqref{18}. 

\emph{Therefore, they were the first to prove that the closed system using fully nonlinear MEP is symmetric hyperbolic if we use the Lagrange multipliers as field variables.}

Although the book did not address the integrability of the moments, their approach was fully nonlinear, which they termed \emph{Molecular Extended Thermodynamics}. They noted the inherent challenge in the nonlinear closure. On p. 171, they wrote regarding the MEP and RET approaches:  
\begin{quote}
    Both theories have one difficulty in common: The inversion of the transformation $\bm{\Lambda} = \bm{\Lambda}(\mathbf{u})$. Indeed, inspection of (1.10), shows that neither $\mathbf{u} = \mathbf{u}(\bm{\Lambda})$ is easy to calculate, nor is that relation easy to invert. In this situation we are forced to linearize just as in extended thermodynamics.
\end{quote}

Consequently, they examined processes near equilibrium and linearized the distribution function using Kogan’s approximation \eqref{appro}. As concrete examples of closure derivations, they explored expansions near equilibrium, deriving maximized nonequilibrium distribution functions for the 20-moment case. They also developed linear field equations for the 14, 20, 21, 26, and 35-moment cases. The 14-moment field equations correspond to Kremer’s phenomenological equations \cite{kremer}.  

\emph{M\"uller and Ruggeri recognized that this linearization sacrifices the elegant property of symmetric hyperbolicity far from equilibrium. They were the first to define, on p. 146, Section 3.2 in \cite{ET}, the so-called "hyperbolicity region" in the space of nonequilibrium variables for which inside the domain the system is hyperbolic, calculating it explicitly for the 13-field case.} This concept was later revisited by other authors. Notably, Brini and Ruggeri \cite{Brini_Ruggeri} demonstrated that considering higher-order polynomial expansions in \eqref{appro} slightly enlarges the hyperbolicity domain.  

Despite the limitations of the hyperbolicity region, the book \cite{ET} highlights important results from Weiss \cite{weiss}, which demonstrated that, with a sufficiently large number of moments, experimental data align closely with RET predictions. This alignment was particularly evident in high-frequency sound wave analysis and light scattering problems. These findings represented a significant success for the theory and contributed to the release of the book's expanded second edition in 1998, where the title was updated from Extended Thermodynamics to Rational Extended Thermodynamics \cite{RET}.

 \subsection{Levermore’s Closure  (1996)}\label{lever}
Three years after the publication of the first edition of M\"uller and Ruggeri's book, Levermore published a paper in 1996 \cite{levermore} that rigorously reformulates the mathematical properties of the MEP, which had previously been noted by Kogan. Unfortunately, Levermore appears to overlook Kogan’s earlier results, which are cited in Dreyer's paper.

A novel aspect of Levermore’s analysis is his explicit focus on the concept of \emph{admissibility} and the emphasis he places on its foundational significance. His so-called "Condition III" closely mirrors Kogan's earlier observation that the integrals involved are not always convergent unless the closure number  
$N$ is even, making this a necessary condition (see Section \ref{koganino}).

Levermore critiques the work of Dreyer and M\"uller-Ruggeri, writing:
\begin{quote}
    Some recent works have employed exponentially based closures of the
form (4.10). Dreyer did so within the context of extended thermodynamics. However, he treated the exponential formally, never
imposing a condition like (III), and proceeded to retain only the quadratic
terms in the exponent while expanding the rest as a polynomial. Both the
entropy and hyperbolic structure are generally lost in the resulting moment
equations.
\end{quote} 
However, this characterization seems to miss critical aspects of Dreyer's contributions. Dreyer explicitly enforces the even-degree polynomial constraint to ensure admissibility and assumes the non-negativity of the distribution function as a fundamental premise. 
Levermore's critiques for linear expansions are confined to the appendix of Dreyer's paper. In the M\"uller-Ruggeri book, it is explained that they serve solely as a tool for inverting the Lagrange multipliers in terms of the physical variables. 

Additionally, contrary to what is stated in the literature, the symmetric form of Godunov-type equations, assuming the integrability of moments, was first explicitly demonstrated in the context of fully nonlinear closure in \cite{ET}.
The fact that hyperbolicity is now restricted to a neighborhood of the equilibrium state was also noted before in \cite{ET}, where the concept of the hyperbolicity region was defined, as previously discussed in Section \ref{MURU}.

From the above discussion, the truly novel aspect of Levermore’s paper, regarding the MEP, in our view, lies in its rigorous mathematical treatment and in the fully nonlinear closure of the ten-moment system. He considered the following $10$ independent fields:  $\left(F=\rho, F_i=\rho v_i, F_{ij}= \rho v_i v_j + p_{ij}\right)$, where $\mathbf{p} \equiv (p_{ij})$ is the pressure tensor (the stress tensor with a sign change) of which trace part is related to  the pressure $p_{ll}=3p$:
\begin{align*}
    p_{ij} = m \int_{\R^3}  f\, C_{i}C_j  \, d\boldsymbol{\xi}, \qquad C_i =  \xi_i - v_i.
\end{align*}
In this case, the distribution function without expansion is explicitly obtained as the following Gaussian form:
\begin{align}
	f_{10} = \frac{\rho }{m\left(2\pi\right)^{3/2} \left[\det \left(\mathbf{p}/\rho\right)\right]^{1/2} } \exp \left\{-\frac{1}{2}  \left(\frac{{\bf p}}{\rho}\right)^{-1}_{ij} C_iC_j \right\}.
    \label{f10}
\end{align}

\subsection{Boillat and Ruggeri’s Analysis of Convergence for $N$ Large and General Entropy Functional (1997)}
Boillat and Ruggeri, in a paper in 1997 \cite{Boillat-Ruggeri}, presented new results on this subject that can be summarized in three parts:  
\begin{itemize}  
    \item They first considered the MEP closure with a generic entropy functional\footnote{In the paper, the authors used $h$ as the entropy, changed in sign as is customary in the mathematical community, which prefers defining convex entropy functions instead of concave ones. Consequently, the signs of the Lagrange multipliers and the main field are opposite to those presented here.}  
    \begin{equation}  \label{hpsi}
    h =\int_{\mathbb{R}^{3}}  \psi (f) d\boldsymbol{\xi} ,
\end{equation}  
    and proved that the MEP closure coincides with the requirement that the truncated moment system satisfies an entropy principle and has a symmetric Godunov form.  
  
    \item They studied the convergence of the distribution function that maximizes the entropy as  $N \rightarrow \infty$.  
  
    \item They demonstrated the existence of a lower bound for the maximal characteristic velocity in equilibrium of the hyperbolic system, showing not only that it increases with $N$ but that it is unbounded in the limit as $N$ tends to infinity.  
\end{itemize}  
\noindent
In the present case  the functional $\mathcal{L}$:
\begin{equation*}
\mathcal{L}= \int_{\mathbb{R}^{3}}  \psi (f) d\boldsymbol{\xi} + \Lambda _{A}\left( F_A-m \int \xi_Af  d\boldsymbol{\xi} \right). 
\end{equation*}%
We impose the extremum  condition, and we have
\begin{equation}\label{psioo}
\frac{d\psi(f) }{df}=m  \chi ,
\end{equation}%
hence it follows that $f(t,\mathbf{x},\bm{\xi})$ is a 
function of a single variable $f\equiv f(\chi)$, with  $\chi = \Lambda _{A}(t,\mathbf{x})\xi_A$. Then \eqref{psioo} give:
\begin{equation*}
\psi (f)=m\left(\chi \F^\prime(\chi)- \F(\chi) \right),   
\end{equation*}%
where $\F$ is the partition function such that $\F^\prime(\chi) = f(\chi)$ and a prime on a quantity indicates the derivative with respect $\chi$. As the extremum needs to be a maximum, we have the inequality $\F^{\prime\prime}<0$.

Taking into account \eqref{4pot}, it is easy to verify that 
 potentials \( h' \) and \( h'^i \) are expressed as:
\begin{align} \label{h'BR}
    h' = \int_{\R^3} \F(\chi)d\boldsymbol{\xi}, \quad h'^i = \int_{\R^3} \xi_i \,\F(\chi)d\boldsymbol{\xi}.
\end{align}
Therefore, if the truncated system is compatible with an entropy principle \eqref{EP}, then \eqref{potenziali} are true, and it is simple to prove that the Lagrange multipliers field  $\bm{\Lambda} \equiv(\Lambda_A)$  coincide with the main field $\mathbf{u}^\prime$ and the two closure of MEP and RET at molecular level are equal also for a generic case.

Inserting \eqref{h'BR} into \eqref{18}, we obtain the closed system for any $N$ is symmetric hyperbolic on the form:
\begin{align*}
    \left(m^2 \int_{\R^3} \F'' \xi_A \xi_B d\boldsymbol{\xi}\right) \partial_t \Lambda^B
    + \left(m^2 \int_{\R^3} \F'' \xi_i \xi_A \xi_B d\boldsymbol{\xi}\right)  \partial_i \Lambda^B = P_A.
\end{align*}
Concerning the convergence for large $N$ and the lower bound for the maximum characteristic velocity, Boillat and Ruggeri had the  idea to map the $(k+1)(k+2)/2$ components of the main field (Lagrange multipliers in the MEP) of order $k$:
\begin{equation*}
{u_{i_{1}i_{2}\dots i_{k}}^{\prime }},\qquad i_{1}\leq i_{2}\leq \dots \leq
i_{k}
\end{equation*}%
in the corresponding variables:
\begin{equation*}
{u_{pqr}^{\prime }},\qquad p+q+r=k.
\end{equation*}%
In fact if we collect all powers of \( \xi_1, \xi_2, \xi_3 \) in \( \chi \), as expressed earlier by \eqref{bravoKogan}$_2$, it can be rewritten using this notation as:
\begin{equation*}
\chi =\sum_{p,q,r}{{u_{pqr}^{\prime }}{\xi_{1}^{p}}{\xi_{2}^{q}}{\xi_{3}^{r}}}%
,\qquad 0\leq p+q+r\leq N.
\end{equation*}
Then, they proved in \cite{Boillat-Ruggeri} the theorem that asserts that, 
for any $N$, we have the following  lower bound condition for the maximum characteristic velocity evaluated in equilibrium, $\lambda _{\max }$:
\begin{equation}
\frac{\lambda _{\max }}{c_0}\geq \sqrt{\frac{6}{5}\left( N-\frac{1}{2}%
\right) },   \label{lowerb}
\end{equation}%
where $c_0$ is the sound velocity. Therefore, $\lambda _{\max }$ increases with $N$ (as was noticed numerically by Weiss \cite{weiss}) and becomes
unbounded when $N\rightarrow \infty $.

We discussed before 
the problem of the convergence of the moments, and according to Kogan, the index of truncation $N$ must be even. 

Moreover, if the conjecture --- the distribution function $f_N$, when $N \rightarrow \infty$, tends to the distribution function $f$ that satisfies the Boltzmann equation --- is true, we need another convergence requirement.

Since
\[
\left| \sum_{p,q,r} u^\prime_{pqr} \xi_1^p \xi_2^q \xi_3^r \right| \leq a_N |\bm{\xi}|^N
\]
with
\[
a_N =\max_{|\mathbf{t}|=1} \nu_N(\mathbf{t}), \qquad \nu_N(\mathbf{t})= \sum_{p,q,r} u^\prime_{pqr} t_1^p t_2^q t_3^r, 
\]
the series is absolutely convergent, when $p + q + r = N \rightarrow \infty$, for any $\bm{\xi}$ provided that
\[
u^\prime_{pqr} \rightarrow 0, \qquad \frac{a_{N+1}}{a_N} \rightarrow 0.
\]
Hence, the components of the main field become smaller and smaller when $N$ increases.  This justifies the truncation of the system. On the other hand, when $N$ is finite, the integrals of moments must also be convergent (Levermore's admissibility discussed in Section \ref{lever}). When $|\bm{\xi}|$ is large, $\chi \simeq |\bm{\xi}|^N \nu_N$.  Therefore, it is easy to see, by using the spherical coordinates, that the integrals of moments converge provided that $\nu_N(\mathbf{t}) < 0$ for any unit vector $\mathbf{t}$. But, since
$\nu_N(-\mathbf{t}) = {(- 1)}^N \nu_N(\mathbf{t})$, 
we can conclude that $N$ must be even and $\max_{|\mathbf{t}|=1} \nu_N(\mathbf{t}) < 0$, and we recover in another way the Kogan's observation and the Levermore's admissibility.

  \subsection{Another  Problematic of MEP  Closure: Junk paper (1998) }\label{Junkino}
Unfortunately, the elegant form derived in the fully nonlinear case has significant challenges. In addition to the previously mentioned issues—namely, concerns about convergence and the difficulty of inverting the mapping between Lagrange multipliers and physical variables—a more fundamental problem was brought to light by Junk \cite{Junk1,Junk2}. Specifically, the complete closure achieved through the MEP introduces a critical limitation: certain physically admissible states correspond to singularities. In other words, for these states, the mapping between Lagrange multipliers and physical variables cannot be inverted.

However, when studying nonequilibrium phenomena near a local equilibrium state, this issue, often referred to as the Junk problem, does not arise. Consequently, for many practical problems, the approximate distribution function proposed by Kogan, as given in \eqref{appro}, is employed despite its associated challenges in defining hyperbolicity regions. Nevertheless, as discussed in Section \ref{MURU}, the theory remains in strong agreement with experimental results.

\smallskip

We conclude this section by citing two other highly interesting papers by Dreyer. The first, co-authored with Kunik and published in 1998 \cite{Dreyer-Kunik}, considers a modification of the MEP to study two extreme cases in kinetic theory: the dominance of particle interactions and free flight. The second, co-authored with Herrmann and Kunik \cite{D-H-K} in 2004, explores another revision of the MEP and discusses Junk's problem in the context of the Boltzmann–Peierls equation, which governs heat conduction in crystalline solids.

   \subsection{Extensions of RET (2011) and MEP (2013) to   Rarefied Polyatomic Gases}
   For a long time, there have been several attempts to generalize both RET and Kinetic Theory from the case of monatomic gases to the more interesting case of polyatomic gases.
Regarding RET, a new approach for rarefied polyatomic gases was developed by Arima, Taniguchi, Ruggeri, and Sugiyama \cite{Arima-2011}.
This theory adopts two parallel hierarchies (binary hierarchy) for the independent fields: mass density, velocity, internal energy, shear stress, dynamic pressure, and heat flux. One hierarchy consists of balance equations for the mass density $F$, momentum density $F_i$, and momentum flux $F_{i_1i_2}$ (\textit{momentum-like} hierarchy), while the other consists of balance equations for twice  the energy density $G_{ll}$ and twice the energy flux $G_{lli}$   (\textit{energy-like} hierarchy);
\begin{align}
\begin{split}
 &\partial_t F + \partial_i F_i = 0,\\
 &\partial_t F_{i_1} + \partial_i F_{ii_1} = 0, \\
 & \partial_t F_{i_1 i_2} + \partial_i F_{ii_1 i_2} = P_{i_1 i_2}, \ \ \ \ \ \ \  \partial_t G_{ll} + \partial_i G_{lli}=0,\\
 &\ \hspace{4.0 cm} \partial_t G_{lli_i} + \partial_i G_{llii_1} =Q_{lli_1},
\end{split}\label{balanceFG}
\end{align}
where $F_{ii_1i_2}$ is the flux of $F_{i_1i_2}$, $G_{llii_1}$ is the flux of $G_{lli_1}$,  
and $P_{i_1 i_2}$ and $Q_{lli_1}$ are the production term associated with $F_{i_1i_2}$ and $G_{lli}$, respectively.
These hierarchies cannot merge with each other, in contrast to the case of rarefied monatomic gases, because the specific internal energy (the intrinsic part of the energy density) is no longer simply related to the pressure (an intrinsic part of the momentum flux).  Moreover, for polyatomic gases, the equations are 14 because among the unknowns there is also the nonequilibrium  dynamical  pressure \( \Pi \), which  is identically zero for  monatomic case.

Regarding the kinetic counterpart, a crucial step in developing the theory of rarefied polyatomic gases was made by Borgnakke and Larsen \cite{Borgnakke-1975}. In their model, the distribution function is assumed to depend on an additional continuous variable representing the energy of the internal modes of a molecule. This allows the model to account for the exchange of energy (beyond translational energy) during binary collisions. Initially, this model was used for Monte Carlo simulations of polyatomic gases and was later applied to derive the generalized Boltzmann equation by Bourgat, Desvillettes, Le Tallec, and Perthame \cite{Bourgat-1994}.  

As a result of introducing an additional parameter \(I\) that has the meaning of microscopic internal energy, the velocity distribution function \(f(t,\mathbf{x},\boldsymbol{\xi},I)\) is defined over an extended domain \([0,\infty) \times \mathbb{R}^3 \times \mathbb{R}^3 \times [0,\infty)\). The quantity $f(t, \mathbf{x}, \boldsymbol{\xi}, I) \varphi(I) \, d\xx \, d\boldsymbol{\xi} \, dI$ represents the {number} of molecules in the 7-dimensional phase space around a point $(\xx, \boldsymbol{\xi}, I)$ at time $t$. Its rate of change is governed by the Boltzmann equation, which retains the same form as that for monatomic gases (\ref{Boltzmann}), but the collision integral \(Q(f)\) incorporates the effects of internal degrees of freedom through the collisional cross-section.  

Pavi\'c, Ruggeri, and Simi\'c \cite{Pavic-2013} demonstrated that the structure of \eqref{balanceFG} can be derived from the Boltzmann equation by identifying the following moments:
\begin{align*}
  & \left( \begin{array}{c}
      F \\
      F_{i_{1}} \\
      F_{i_1 i_2} \\
    \end{array}\right) = \int_{\mathbb{R}^{3}}
    \int_{0}^{\infty} m
    \left(%
    \begin{array}{c}
      1 \\
      \xi_{i_{1}} \\
      \xi_{i_{1}} \xi_{i_{2}} \\
    \end{array}%
    \right)
    f(t,\mathbf{x},\boldsymbol{\xi},I) \,
    \varphi(I) \, dI \, d\boldsymbol{\xi},
 \nonumber \\
  \\
  & \left(\begin{array}{c}
      G_{ll} \\
      G_{lli_1} \\
    \end{array}\right) = \int_{\mathbb{R}^{3}}
    \int_{0}^{\infty} m
    \left(%
    \begin{array}{c}
       \xi^{2} +2 \frac{I}{m} \\
      \left(  \xi^{2} +2 \frac{I}{m}
        \right) \xi_{i_{1}} \\
    \end{array}%
    \right)
    f(t,\mathbf{x},\boldsymbol{\xi},I) \,
    \varphi(I) \, dI \, d\boldsymbol{\xi}.
  \nonumber
\end{align*}
The weighting function $\varphi(I)$ is determined in 
such a way that it recovers the caloric equation of state in equilibrium for
polyatomic gases. 
Then,  they used the MEP, in the case of $14$ moments, to yield the appropriate macroscopic balance laws under the assumptions of processes near equilibrium. The structure of the distribution function that satisfies the MEP is similar to the one of Kogan \eqref{bravoKogan} with a more complex $\chi$:
\begin{equation*}
    \chi= \Lambda + \Lambda_i \xi_i + \Lambda_{ij}\xi_i \xi_j  + \Omega \left(\xi^{2} +2 \frac{I}{m}\right) + \Omega_i\left(\xi^{2} +2 \frac{I}{m}\right)\xi_i
\end{equation*}
that takes into account the Lagrange multipliers $\Lambda$'s of the $F$'s hierarchy and the $\Omega$'s of the $G$ hierarchy. They showed complete agreement with the RET procedure given in \cite{Arima-2011}.
Later, using the MEP, Ruggeri studied the closure in the most complicated case of non-polytropic gas, in which the internal energy is a nonlinear function of the temperature \cite{Ruggeri_Wascom19}\footnote{In ref. \cite{Pavic-2013}, there are typos that were correct in \cite{Ruggeri_Wascom19} and in the book \cite{newbook}.}.

Therefore, even for rarefied polyatomic gases, the closure procedures of RET and MEP yield the same result for any non-polytropic gas. More details, including recent results on polyatomic gases, can be found in the book \cite{newbook} and the references therein.

Although the theory is developed in the neighborhood of equilibrium, one of its most remarkable successes is that, unlike in the case of monatomic gases, for certain gases with large bulk viscosity, a small number of moments (6 or 14) are sufficient to describe the experimental data on shock structure  \cite{Taniguchi1,Taniguchi2} and of sound waves \cite{Sound}. Furthermore, numerical simulations of shock waves conducted by Kosuge et al. \cite{Kosuge1,Kosuge2,Kosuge3} using the Boltzmann equation for polyatomic gases are indistinguishable from those obtained using the RET theory.

Finally, we remark that the model for monatomic gases can be derived from that of polyatomic gases as its singular limit (see \cite{newbook} and references therein).

\subsection{Fully Nonlinear MEP Closure for Polyatomic Gases}
In Section \ref{Junkino}, we discussed the challenges of the nonlinear MEP closure, which was criticized by Henning Struchtrup in his book \cite{henning}. In Section 6.6.2, on page 106, he wrote:
\begin{quote}
The elegant mathematical formulation of the theory of maximization of entropy is far outweighed by the problems arising with the method, and its use cannot be recommended.
\end{quote}

While this criticism raises important points, this statement is not entirely acceptable. In the case of a theory near equilibrium, MEP  still provides a valid tool for closing the system. The fact that hyperbolicity is not global but only in a specific domain is not unusual in nonlinear physical theories. Indeed, global hyperbolicity is an excessively stringent requirement, and only ideal cases (e.g., Euler fluid with the ideal gas assumption) satisfy this requirement for all possible values of the field variables. The lack of hyperbolicity could also be associated with physical effects (phase transition, instability, etc.), as, for example, in van der Waals fluids, in Born–Infeld nonlinear electrodynamics, in nonlinear elasticity, and in the Boltzmann–Vlasov equation (see \cite{Brini_Ruggeri} and the references therein). The system, within the hyperbolic domain, has all the beneficial properties of qualitative analysis due to its nature as a symmetric hyperbolic system that satisfies the so-called K-condition. In particular, there exist global smooth solutions for initial data sufficiently small (for further clarification, interested readers can refer to \cite{newbook} and references therein).

Moreover, there are examples where nonlinear closure is possible, and it is also feasible to invert the Lagrange multipliers as functions of the physical variables. Beyond the discussed case of the 10-moment closure by Levermore mentioned before in Section \ref{lever}, there are, in the polyatomic case, some examples of nonlinear closure that we now mention here.

\subsubsection{MEP of Polyatomic Gas with Six Moments}
The RET with ${14}$ fields, \eqref{balanceFG}, provides a satisfactory phenomenological model, but its differential system is rather complex and is valid only near equilibrium. For this reason, a simplified theory with $6$ fields (RET$_6$) \cite{ET6,ET6nonlinear}—the mass density $\rho$, the velocity $v_i$, the temperature $T$, and the dynamic pressure  $\Pi$—is constructed.

This simplified theory preserves the main physical properties of the more complex RET theory with $14$ variables, particularly when the bulk viscosity plays a more important role than the shear viscosity and heat conductivity. This situation is observed in many gases, such as rarefied hydrogen gas and carbon dioxide gas at certain temperature ranges. In this case, in \eqref{balanceFG}, we consider only the trace equation of $F_{i_1i_2}$ and disregard the equation for the heat flux for $G_{lli_1}$.

The nonlinear version of RET was developed in \cite{ET6nonlinear}, while the nonlinear closure via MEP for this model was provided in \cite{R-B-S} for polytropic gases and in \cite{Bisi_Ruggeri_Spiga} for non-polytropic ones.
In the present case, the $6$ moments are:
\begin{equation}\label{F6}
  \left(%
    \begin{array}{c}
    F \\
   F_{i} \\
   F_{ll}\\
    \end{array}%
    \right) =
  \left(%
  \begin{array}{c}
  \rho \\
  \rho v_{i} \\
  \rho v^2 + 3(p+\Pi) \\
  \end{array}%
  \right)=
  \int_{\mathbb{R}^{3}} \int_{0}^{\infty}
        m \left(%
          \begin{array}{c}
          1 \\
          \xi_{i} \\
         \xi^2 \\
          \end{array}%
        \right)
        f \varphi(I) \, dI \,
        d\bm{\xi} ,
\end{equation}
and
\begin{equation*} 
G_{ll}=\rho v^2 + 2 \rho \varepsilon= \int_{\mathbb{R}^{3}} \int_{0}^{\infty} m \left(\xi^2+2 \frac{I}{m}\right)f \varphi(I)  \, dI \
         d\bm{\xi}.
\end{equation*}
Here, the internal energy $\varepsilon$ is the sum of the kinetic part due to translation and the internal motion due to rotation and vibration of a molecule: 
\begin{align}
& \varepsilon= \varepsilon^K + \varepsilon^I, \quad \text{with} \nonumber \\
& \varepsilon^K= \int_{\mathbb{R}^{3}} \int_{0}^{\infty} \frac{1}{2} m C^2 f \varphi(I)  dI
d\bm{\xi}, \quad \varepsilon^I= \int_{\mathbb{R}^{3}} \int_{0}^{\infty} I f \varphi(I)  dI
d\bm{\xi}. \label{interna}
\end{align} 
In equilibrium, we have the caloric equations of state 
\begin{align}
    \varepsilon^K_E = \frac{3}{2}\frac{k_B}{m}T, \quad \varepsilon^I_E = \varepsilon^I_E(T),
\end{align}
where the index $E$ indicates the same quantities in equilibrium.

In the present simple case, it is possible to invert the Lagrange multipliers in terms of physical variables far from equilibrium. In \cite{R-B-S}, the nonequilibrium distribution function that satisfies the MEP was determined as follows:
\begin{equation*} 
f_6 = \frac{\rho}{m (2\pi)^{3/2} A(T^I) }\left(\frac{\rho}{p+\Pi}\right)^{3/2}  \exp \left( - \frac{mC^2}{2 k_B T} \frac{\rho}{p+\Pi} - \frac{I}{k_B T^I}\right),  
\end{equation*} 
where the nonequilibrium temperature $T^I$ is defined implicitly through the caloric equation of state as follows:
\begin{align*}
    \varepsilon^I_E(T^I) = \varepsilon^I , 
\end{align*}
where $\varepsilon^I$ is given by \eqref{interna}$_2$.
While considering $\varepsilon=\varepsilon^K + \varepsilon^I = \varepsilon^K_E + \varepsilon^I_E$ and using $2\rho\varepsilon^K_E = 3p$ and $2\rho\varepsilon^K = 3(p+\Pi)$, where the latter relation is derived from \eqref{F6} and \eqref{interna}$_1$, $T^I$ is related to the dynamic pressure $\Pi$ through the relation: 
\begin{equation*} 
\frac{\varepsilon_E^I (T)-\varepsilon_E^I (T^I)}{\varepsilon_E^K (T)}=\frac{\Pi}{p}.
\end{equation*} 
The normalization factor $A\left(T^I\right)$ is given by 
\begin{equation*} 
A\left(T^I\right) = A_0 \exp\left(\frac{m}{k_B}\int_{T_*}^{T^I} \frac{\varepsilon_E^I({x} )}{{x} ^2} d{x} \right),
\end{equation*} 
where $A_0$ is an inessential constant and $T_*$ is a reference temperature.
All the moments are convergent, and the bounded solutions satisfy the inequalities: 
\begin{equation*} 
-1<\frac{\Pi}{p} < \frac{\varepsilon_E^I (T)}{\varepsilon_E^K (T)}. 
\end{equation*}

\subsubsection{Ellipsoidal Gaussian Distribution: MEP for Polyatomic Non-polytropic Gas  with Eleven Moments}
By considering up to the second-order moments, similar to the 6-moment case, but including all components, a system formed by the 11-moment; $(F,F_i,F_{ij},G_{ll})$, where $F_{ij}=\rho v_i v_j + p_{ij}$  can be constructed. Here, the pressure tensor $\mathbf{p} \equiv (p_{ij})$ for polyatomic gases is defined by
\begin{align*}
    p_{ij} = m \inta  f\, C_{i}C_j  \varphi(I)  dI d\bm{\xi}.
\end{align*}
being $p_{ll} =3( p+\Pi)$.

As a result of MEP, the distribution function of the 11-moment system is given by \cite{lastESBGK}
\begin{align*}
	f = \frac{\rho }{m\left(2\pi\right)^{3/2} \left[\det \left({\bf p}/\rho\right)\right]^{1/2} A\left(T^I\right)} \exp \left\{-\frac{1}{2} \left(\frac{{\bf p}}{\rho}\right)^{-1}_{ij} C_i C_j -\frac{I}{k_B T^I}\right\}.
\end{align*}
This distribution function is a natural extension of \eqref{f10} to polyatomic gases.

\subsection{MEP Incorporating Relaxation Processes of Molecular Rotational and Vibrational Degrees of Freedom}

As observed in RET$_6$, the RET for polyatomic gases described so far treats the internal energy $\varepsilon^I$ as a single entity representing the molecular internal degrees of freedom. However, for gases such as carbon dioxide gases, where slow relaxation of molecular vibrational modes is observed, it becomes necessary to separately describe the nonequilibrium relaxation processes of rotational and vibrational modes. 

To address this, in \cite{ET7}, a kinetic model of gas molecules was proposed, in which the microscopic energy parameter $I$ is divided into parameters representing vibrational and rotational modes. 
In this case, since $\varepsilon^I$ is separated into rotational and vibrational energies, a 7-field theory was constructed as a result of the nonlinear MEP corresponding to RET$_6$. 

Furthermore, although using an expansion near equilibrium, the 15-field MEP corresponding to a 14-field theory has also been studied \cite{ET15}.


\section{MEP in Relativistic Kinetic Theory of Gases} \label{sec:rel}
In 1987, the emergence of the new approach of Extended Thermodynamics garnered significant interest. The ISIMM (International Society for the Interaction of Mechanics and Mathematics) invited M\"uller and Ruggeri to organize a symposium, which took place in Bologna. The proceedings were published by a local publishing house, Pitagora Editrice (Bologna).

In this volume, there is a contribution by Dreyer \cite{Dreyer2} on the first use of the MEP in the context of a relativistic fluid. This result was never subsequently published in a journal and thus fell into relative obscurity. One of us, T.R., despite being an editor of the volume, had completely forgotten about Dreyer's work. In fact, even when applying the MEP method to the case of a relativistic polyatomic gas \cite{pennisi} or in the recent book co-authored with Sugiyama \cite{newbook}, Dreyer’s results were never cited.
Some time ago in 2022, T.R. received an email from  Stefano Boccelli, working at that time as a postdoc in Canada under the supervision of James McDonald, requesting a copy of Dreyer's paper. This gave us the opportunity to revisit it, which we briefly summarize here.

\subsection{Relativistic Kinetic Theory and Moment Equations}
In relativity for Minkowski space, the Boltzmann--Chernikov relativistic equation read  \cite{C-K,Chernikov,Synge}:
  \begin{equation}\label{BoltzR}
  p^\alpha \partial_\alpha f = Q,
  \end{equation}
 in which the distribution function  $f$ depends on  $(x^\alpha,  p^\beta)$, where $x^\alpha$ are the space-time coordinates, $p^\alpha$ is the four-momentum, $\partial_{\alpha} = \partial/\partial x^\alpha$,  $Q$ is the collisional term, and $\alpha, \beta =0,1,2,3$. 
 For monatomic gases, the relativistic moment equations associated with \eqref{BoltzR}, truncated at tensorial index $N+1$ are:
  \begin{equation}\label{Relmomentseq}
  \partial_\alpha A^{\alpha \alpha_1 \cdots \alpha_n  } =  I^{  \alpha_1 \cdots \alpha_n   }
  \quad \mbox{with} \quad n=0 \, , \,\cdots \, , \,  N 
  \end{equation}
  with  
  \begin{align}\label{RelmomentMono}
  \begin{split}
  &A^{\alpha \alpha_1 \cdots \alpha_n  } = \frac{c}{m^{n-1}} \int_{\R^{3}}
  f  \,  p^\alpha p^{\alpha_1} \cdots p^{\alpha_n}  \, \, d \boldsymbol{P}, \\
  &I^{\alpha_1 \cdots \alpha_n  } = \frac{c}{m^{n-1}} \int_{\R^{3}}
  Q  \,   p^{\alpha_1} \cdots p^{\alpha_n}  \, \, d \boldsymbol{P},
  \end{split}
  \end{align}
  where $c$ denotes the light velocity, $m$ is the particle mass in the rest frame,
 and 
  \begin{equation*}
  d \boldsymbol{P} =  \frac{dp^1 \, dp^2 \,
  	dp^3}{p^0} .
  \end{equation*}
If $n=0$, the tensor reduces to $A^\alpha$; moreover, the production tensor in the right side of \eqref{Relmomentseq} is zero for $n=0,1$  because the first $5$ equations represent the conservation laws of the particle number and of the energy-momentum, respectively.

When $N=1$, we have the relativistic Euler system
  \begin{equation*}
     \partial_\alpha A^{\alpha } = 0, \quad \partial_\alpha A^{\alpha \beta} =0, 
     \end{equation*}
     where, also in the following,  $A^\alpha \equiv V^\alpha$ and $A^{\alpha\beta} \equiv T^{\alpha\beta}$ have the physical meaning, respectively, of the particle number vector and the energy-momentum tensor;
     \begin{align}
         V^\alpha =  \rho U^\alpha, \quad T^{\alpha\beta} = p h^{\alpha \beta} + \frac{e}{c^2} U^{\alpha } U^{\beta}, \label{VTE}
     \end{align}
     where $\rho=n m$, with $n$ being the particle number density, $U^\alpha$ is the four-velocity $(U^\alpha U_\alpha = c^2)$, $p$ is the pressure, $e$ is the energy, and $h^{\alpha\beta}$ is the projector:
     \begin{align*}
         h^{\alpha\beta} = - g^{\alpha\beta} + \frac{1}{c^2} U^\alpha U^\beta,
     \end{align*}
     being  $g^{\alpha \beta}= \text{diag}(1 \, , \, -1 \, , \, -1 \,, \, -1)$  the metric tensor. 
In this  equilibrium case, the distribution function reduces to the J\"uttner distribution function that has the following expression for any kind of gas:
\begin{align}
    f_J = \frac{Y}{\exp\left(\frac{m}{k_B} \alpha + \frac{U_\alpha p^\alpha}{k_B T}\right) +s},
\end{align}
where $\alpha$ denotes the fugacity and is expressed as $\alpha = - g/T$ being $g$ the chemical potential.

When $N=2$, we have the system:
  \begin{equation}
   \partial_\alpha A^{\alpha } = 0, \quad \partial_\alpha A^{\alpha \beta} =0, \quad  \partial_\alpha A^{\alpha \beta \gamma} =  I^{  \beta \gamma  }, \label{Annals}
   \end{equation}
   that is formed of $14$ equations taking into account that $A^{\alpha\beta}_{\,\,\,\,\,\,\,\beta} = m^2 c^2 A^\alpha$ and $I^\beta_{\,\,\,\beta}=0$.
   In this case, the particle number vector and the energy-momentum tensor are decomposed as
\begin{align*} 
    V^\alpha =\rho  U^\alpha \, , \quad  T^{\alpha \beta} = \frac{e}{c^2} \,  U^{\alpha } U^\beta + \, \left(p \, + \, \Pi\right)
    h^{\alpha \beta} + \frac{1}{c^2} ( U^\alpha  q^\beta +U^\beta  q^\alpha)+   t^{\langle\alpha \beta\rangle_3} \, ,
\end{align*}
with $14$ physical variables;  $\rho$, $T$, $U^\alpha$, $\Pi$, $q^\alpha$, $t^{\langle\alpha \beta\rangle_3}$, where $\Pi$ is the dynamic pressure, $q^\alpha= -h^\alpha_\mu U_\nu T^{\mu \nu}$ is the heat flux, and $t^{\langle\alpha \beta\rangle_3} = T^{\mu\nu} \left(h^\alpha_\mu h^\beta_\nu - \frac{1}{3}h^{\alpha\beta}h_{\mu\nu}\right)$ is the deviatoric shear viscous stress tensor.  We also recall the constraints:
\begin{equation*}
U^\alpha U_\alpha = c^2, \quad q^\alpha U_\alpha = 0, \quad t^{\langle\alpha \beta\rangle_3} U_\alpha = 0, \quad t^{\langle\alpha}_{\,\,\,\,\,\ \alpha \rangle_3} =0.
\end{equation*}

As it is well known, the pioneering papers by M\"uller \cite{mueller1} and Israel \cite{Isr} are the first tentative to obtain a causal relativistic phenomenological theory with a system of equations of a hyperbolic type such that the wave speeds are finite consistently with the relativity principle. For the same reason as the classical case, Liu, M\"uller and Ruggeri (LMR) \cite{LMR} (see also \cite{RET}) explored the possibility of having a new theory that starts with a few natural assumptions and that uses only universal principles. Therefore, they considered the structure of the $14$ moments \eqref{Annals} and closed the system \eqref{Annals} using the RET procedure at the phenomenological level.

\subsection{MEP for Monatomic Relativistic  Gas: Dreyer (1987)}

The closure of MEP of the system \eqref{Annals} was the object of the important and complex paper of Dreyer \cite{Dreyer2}. He required, under the constraints that the temporal parts $V^\alpha U_\alpha$ and $T^{\alpha\beta}U_\beta$  are prescribed, to find  the appropriate distribution function $f\equiv f(x^\alpha,   p^\alpha)$, which maximizes the entropy density $h=h^\alpha U_\alpha$ 
where $h^\alpha$ is the four entropy described by
\begin{align*}
    h^\alpha = - k_B \, c \,  U_\alpha \int_{\R^3} \left\{ \left(s^2 -1 + \ln \frac{f}{\y}\right) + s \frac{\y}{f} \left( 1 - s \frac{f}{\y}\right) \ln \left( 1 -s \frac{f}{\y} \right) \right\}f \, d \boldsymbol{P} \, ,
\end{align*}
which is the relativistic counterpart of \eqref{hgen}.

The distribution function that maximizes the entropy retains the same form as the classical ones given in \eqref{tom}. However, in the relativistic theory, even in near equilibrium, the complexity of Dreyer's paper lies in the inversion between the Lagrange multipliers and the field variables.

Despite the groundbreaking nature of Dreyer’s work \cite{Dreyer2} on the closure for the relativistic 14-moment system for degenerate gases, it has largely been overlooked, as it appears only in lecture notes \cite{Dreyer2}, and the present authors are not aware of it being published elsewhere\footnote{As we cannot provide all details here, readers interested in Dreyer's paper \cite{Dreyer2} may contact one of the present authors, T.R., via email. T.R. will be happy to provide a copy of the relevant pages from the volume containing the paper.}.

Flowing Dreyer let us close the system \eqref{Annals} in which $A^{\alpha\beta\gamma}$ remains to determine in terms of the $14$ physical fields. 
As usual, to obtain a specific expression of the closed field equations, we consider the processes near equilibrium and therefore expand around an equilibrium state as follows \footnote{In Dreyer's paper \cite{Dreyer2}, the case of the case of zero rest-mass is also studied. For simplicity, here we excluded this case.}:
\begin{align}\label{fgenE}
 \begin{split}
 &f \simeq f_J\Big(1-\frac{1}{k_B}\tilde{\chi}\Big), \\
 &\tilde{\chi}   = m \, (\lambda - \lambda_E) \, + \,  (\lambda_{\mu}-\lambda_{\mu_E}) \, p^{\mu}  + \,  \frac{1}{m} \, \lambda_{\mu \nu} \, p^{\mu} p^{\nu}   .
 \end{split}
\end{align}
Inserting the distribution function \eqref{fgenE} into the moments \eqref{RelmomentMono}, we obtain the algebraic system for the Lagrange multipliers in terms of the physical variables. 

The calculations are very cumbersome due to the appearance of the integrals:
\begin{align}
    &J_{\mu,\nu}(\alpha, \gamma) = \int_0^\infty \frac{\sinh^\mu \rho \cosh^\nu \rho }{\exp\left(\frac{m}{k_B}\alpha + \gamma \cosh \rho \right) \mp 1}d\rho.
\end{align}
The $J_{\mu,\nu}$ are functions
of the fugacity $\alpha$ and the relativistic dimensionless coldness $\gamma$ defined by 
\begin{align}
\gamma = \frac{mc^2}{k_BT}. 
\end{align}
First, the expressions for the equilibrium variables were obtained as follows:
\begin{align}
\begin{split}
n = 4 \pi \y m^3 c^3 J_{2,1}, \quad
e = 4 \pi \y m^4 c^5 J_{2,2}, \quad
p = \frac{4}{3} \pi \y m^4 c^5 J_{4,0},
\end{split}
\label{relEval}
\end{align}
then, Dreyer derived the following closure expression for the triple tensor in terms of physical variables:
\begin{align}
A^{\alpha\beta\gamma} = & \; (C_0^1 + C_\pi^1 \Pi) U^\alpha U^\beta U^\gamma 
+ \frac{c^2}{6} \big(nm - C_0^1 - C_\pi^1 \Pi\big) \big(g^{\alpha\beta} U^\gamma + g^{\alpha\gamma} U^\beta + g^{\beta\gamma} U^\alpha \big) \nonumber \\
& + C_0^3 \big(g^{\alpha\beta} q^\gamma + g^{\alpha\gamma} q^\beta + g^{\beta\gamma} q^\alpha \big)
- \frac{6}{c^2} C_0^3 \big(U^\alpha U^\beta q^\gamma + U^\alpha U^\gamma q^\beta + U^\beta U^\gamma q^\alpha \big) \nonumber \\
& + C_0^5 \big(t^{\langle \alpha\beta \rangle_3} U^\gamma + t^{\langle \alpha\gamma \rangle_3} U^\beta + t^{\langle \beta\gamma \rangle_3} U^\alpha \big),
\end{align}
where
\begin{align*}
 &C_0^1 = m n + 8 \pi Y m^4 c^3 J_{4,1}, \quad 
 C_\pi^1 = 6 \frac{1}{c^2}
 \frac{
\left|\begin{array}{ccc}
J'_{2,1} & J'_{2,2} & J'_{2,3} \\
J'_{2,2} & J'_{2,3} & J'_{2,4} \\
J'_{4,1} & J'_{4,2} & J'_{4,3}
\end{array}\right|
}{
\left|\begin{array}{ccc}
J'_{2,1} & J'_{2,2} & J'_{2,3} \\
J'_{2,2} & J'_{2,3} & J'_{2,4} \\
J'_{4,0} & J'_{4,1} & J'_{4,2}
\end{array}\right|
},\ \
 C_0^3 = -\frac{1}{5} 
\frac{
\left|\begin{array}{cc}
J'_{4,0} & J'_{4,1} \\
J'_{4,2} & J'_{4,3}
\end{array}\right|
}{
\left|\begin{array}{cc}
J'_{4,0} & J'_{4,1} \\
J'_{4,1} & J'_{4,2}
\end{array}\right|
}, \\
&C_0^5 =  \frac{J'_{6,1}}{J'_{6,0}},
\end{align*}
with
\begin{align}
& J'_{\mu,\nu} = \frac{\partial J_{\mu,\nu}(\alpha, \gamma)}{\partial \alpha}.
\end{align}
In \cite{Dreyer2}, it is noticed that the integrals $J_{\mu,\nu}$ and $J'_{\mu,\nu}$ are related to  the following integrals:
\begin{align}
I_\nu (\alpha, \gamma) = \int_0^\infty \frac{\cosh \nu \rho}{\exp \left(\frac{m}{k_B} \alpha + \gamma \cosh \rho \right) \mp 1} d\rho,
\label{Jmn}
\end{align}
in particular, the following relations are shown
\begin{align*}
&J_{2,1} = \frac{1}{4}(I_3 - I_1), \quad J_{2,2} = \frac{1}{8}(I_4 - I_0), \quad J_{4,0} = \frac{1}{8}(I_4 - 4I_2 + 3I_0), \\ 
& J_{2,1}' = -\frac{m}{k_B\gamma} I_2,  \quad 
J_{4,1} = \frac{1}{16}(I_5 - 3I_3 + 2I_1), \quad J_{2,2}' = -\frac{m}{4k_B\gamma}(3I_3 + I_1), \\
& J_{2,3}' = -\frac{m}{2k_B\gamma}(I_4 + I_2),  \quad
J_{2,4}' = -\frac{m}{16k_B\gamma}(5I_5 + 9I_3 + 2I_1), \\ 
& J_{4,0}' = -\frac{3m}{4k_B\gamma}(I_3 - I_1), \quad
J_{4,1}' = -\frac{m}{2k_B\gamma}(I_4 - I_2), \\
&J_{4,2}' = -\frac{m}{16k_B\gamma}(5I_5 - 3I_3 - 2I_1), \quad J_{4,3}' = -\frac{3m}{16k_B\gamma}(I_6 - I_2), \\
&J_{6,0}' = -\frac{5m}{16k_B\gamma}(I_5 - 3I_3 + 2I_1), \quad
J_{6,1}' = -\frac{m}{32k_B\gamma}(6I_6 - 16I_4 + 10I_2).
\end{align*}
The evaluation of these integrals depends on the specific properties of the gas. 
The value of $\frac{m\alpha}{k_B \gamma}$  determines the physical regime. 
For instance, the system corresponds to a non-degenerate gas when \( \frac{m\alpha}{k_B} \to \infty \), 
while for strongly degenerate Fermions, it lies within the range \( -\infty < \frac{m\alpha}{k_B \gamma} < -1 \).
In the case of completely degenerate Fermions, the condition is 
$\frac{m\alpha}{k_B \gamma} \to -\infty$.
Finally, for completely degenerate Bosons, the range is given by 
$\frac{m\alpha}{k_B \gamma} = -1$.
Table \ref{table} highlights how the evaluation of \( I_\nu(\alpha, \gamma) \) varies across different gas regimes, 
providing insight into the transition from non-degenerate to strongly degenerate conditions. In the following table is reported the Dreyer results for $I_\nu$:
\begin{table}[htbp]
\renewcommand{\arraystretch}{1.8}
\centering
\begin{tabular}{|l|l|}
\hline
{Gases} & {Expression for $I_\nu(\alpha, \gamma)$} 
\\ \hline \hline

\multicolumn{1}{|p{1.5cm}|}{\raggedright Non-\\degenerate  }
&
$\ds K_\nu(\gamma) \exp\left(-\frac{m\alpha}{k_B}\right)$ 
\\ \hline

\multicolumn{1}{|p{1cm}|}{\raggedright Strongly \\ degenerate \\ (Fermions) } &
\multicolumn{1}{|p{7cm}|}{\raggedright $\ds \frac{\sinh \nu \rho_F}{\nu} (1 +X + \cdots)$ \\ 
\  with \ $\ds X= \frac{\pi^2 \nu}{6 \gamma^2} \frac{\nu \sinh \nu \rho_F \sinh \rho_F - \cosh \nu \rho_F \cosh \rho_F}{\sinh \nu \rho_F \, \sinh^3 \rho_F}$} 
\\ \hline

\multicolumn{1}{|p{1cm}|}{\raggedright Completely \\ degenerate \\ (Fermions)} &
$\ds \frac{1}{\nu} \sinh \nu \rho_F $ 
\\ \hline

\multicolumn{1}{|p{1cm}|}{\raggedright Completely \\ degenerate\\ (Bosons)} &
$\ds \sum_{r=1}^\infty K_\nu(r\gamma) \exp(r\gamma)$ 
\\ \hline

\end{tabular}
\caption{Expressions for $I_\nu(\alpha, \gamma)$ for non-zero rest mass. $\rho_F = \mathrm{arcosh} \left| {\alpha}/{(k_B \gamma)}\right|$ which characterize Fermi degeneracy, 
and $K_\nu(\gamma)$ is the modified Bessel function: $K_\nu (\gamma) = \int_0^{\infty} \cosh(\nu s) \, e^{-\gamma \cosh s} \, ds $.
}
\label{table}
\end{table}

Dreyer also proved that the results obtained using the closure of the MEP are consistent with the phenomenological LMR closure  \cite{LMR} and coincide with the Marle closure \cite{Marle}, which is the relativistic counterpart of the classical Grad method.

\subsection{Boillat and Ruggeri's Extension to General Entropy Functional (1997)}
As in the non-relativistic gas case (see \eqref{hpsi}), the application of the MEP approach to relativistic expectations for a general entropy functional was carried out by  Boillat \& Ruggeri \cite{BRrel1,BRrel2}, and in this case, the authors proved an upper bound limit for the maximum characteristic velocity similar to the corresponding classical one \eqref{lowerb} but as we aspect they proved that for $N\rightarrow \infty$ the limit is the light velocity according with the expectation of relativistic Boltzmann-Chernikov equation given by   Cercignani \cite{Cerc}. Again, this is another indication of the conjecture (never proved) that the $f_N$ obtained by the MEP converges for $N\rightarrow \infty$ to the solution of \eqref{BoltzR}.

\subsection{MEP for Polyatomic Relativistic Gas (2017)}
Pennisi and Ruggeri first constructed a relativistic RET theory for polyatomic gases with \eqref{Relmomentseq}  in the case of $N=2$ \cite{Annals}, and generalized to many moment case \cite{PRJSP}. After that, from a Pennisi idea, a more physically reliable model was presented with the following moments \cite{ACPR}:
\begin{align} \label{relRET}
\begin{split}
& A^{\alpha \alpha_1 \cdots \alpha_n  } = \Big(\frac{1}{mc}\Big)^{2n-1} \int_{\R^{3}}
\int_0^{+\infty} f  \,  p^\alpha p^{\alpha_1} \cdots p^{\alpha_n}  \, \Big( mc^2 +  I \Big)^n\, 
\phi({I}) \, d {I} \, d \boldsymbol{P}  \, , \\
& I^{\alpha_1 \cdots \alpha_n  } = \Big(\frac{1}{mc}\Big)^{2n-1} \int_{\R^{3}}
\int_0^{+\infty} Q  \,  p^{\alpha_1} \cdots p^{\alpha_n}  \, \Big( mc^2 +  I \Big)^n\, 
\phi({I}) \, d {I} \, d \boldsymbol{P}. \\
\end{split}
\end{align}
where the distribution function $f(x^\alpha, p^\beta,{I})$ depends on the extra variable ${I}$, similar to the classical one (see \cite{RS} and references therein), that has the physical meaning of the molecular internal energy of internal modes in order to take into account the exchange of energy due to the rotation and vibration of a molecule, and $\phi(I)$ is the state density of the internal mode. 

In the case of $N=2$, the moments \eqref{relRET} provides $15$ moments. For such a case,   in \cite{ACPR}, the closed field equations near equilibrium were obtained by adopting MEP.

\subsection{Classical Limit of MEP}
As usual, relativistic theories have compactness properties that are lost in the classical limit, and many properties of classical theories can be understood by considering them as the limit of the corresponding relativistic theories. It is noteworthy that in the relativistic case, there is no issue with the integrability of moments, unlike in the classical case. The reason for this is physically evident, as particle velocities are limited by the speed of light. From a mathematical perspective, this can also be demonstrated (see, for instance, \cite{Integr}).

Moreover, by taking the classical limit in the monatomic case and truncating the relativistic moment system at tensorial order \( N \), Pennisi and Ruggeri \cite{PRJSP} proved that, in the classical limit, there exists a precise hierarchy of moments (the only admissible one). In this hierarchy, some indices are free while others are repeated (trace part), such that the closure tensorial index of the classical limit is 
$N_{\text{class}} = 2N$.
This value is an even number and thus represents a necessary condition for the integrability of moments.  
Additionally, not all classical numbers of moments are admissible. Specifically, for a monatomic gas, if the relativistic system is 
\eqref{Relmomentseq}, then the total number of classical moments, \( \mathcal{N} \), is given by the following formula: 
\begin{equation*}
    {\cal{N}} = \frac{1}{6}(N+1)(N+2)(2N+3).
\end{equation*}
This result agrees with the findings of Dreyer and Weiss \cite{DreyerWeiss}, who showed that for \( N = 2 \), the LMR theory converges in the classical limit to Kremer's theory with \( \mathcal{N} = 14 \) moments \cite{kremer}, rather than to Grad's 13-moment theory. A similar result has been obtained for the polyatomic case \cite{ACPRlast,newbook}.

\section{Conclusions}
Our aim here has been to provide a brief chronological account of how the MEP has been applied in nonequilibrium thermodynamics based on kinetic theory. While our narrative is necessarily limited to selected works, we observe that perhaps the most significant contributions to this subject were made, in temporal order, by Kogan, Dreyer, and, later, in the first edition of the book by M\"uller and Ruggeri.  

In particular, Dreyer was not only the first to consider, in the classical context, the case of a general gas—including Fermions and Bosons—but also the first to apply the MEP closure to a relativistic fluid. This challenging work remains largely unknown, as it was published only in the proceedings of a conference by a local publisher. Without diminishing the contributions of Levermore—who deserves credit for rigorously promoting the topic at an international level—the authors modestly believe that the method for closing moment equations should be referred to as the \emph{Kogan-Dreyer method}.  

\begin{acknowledgement}
The work of T. A. was partially supported by JST, PRESTO Grant Number JPMJPR23O1, Japan, and by JSPS KAKENHI Grant Numbers JP22K03912.
The work of T. R. was carried out in the framework of the activities of the Italian National Group for Mathematical Physics [Gruppo Nazionale per la Fisica Matematica (GNFM/INdAM)].
\end{acknowledgement}

\end{document}